# Performing Informetric Analysis on Information Retrieval Test Collections: Preliminary Experiments in the Physics Domain

Tamara Heck[1] and Philipp Schaer[2]

[1] *tamara.heck@uni-duesseldorf.de*
Heinrich-Heine-Universität Düsseldorf, Universitätsstr. 1, 40225 Düsseldorf (Germany)

[2] *philipp.schaer@gesis.org*
GESIS – Leibniz Institute for the Social Sciences, Unter Sachsenhausen 6-8, 50667 Köln (Germany)

**Abstract**
The combination of informetric analysis and information retrieval allows a twofold application. (1) While informetrics analysis is primarily used to gain insights into a scientific domain, it can be used to build recommendation or alternative ranking services. They are usually based on methods like co-occurrence or citation analyses. (2) Information retrieval and its decades-long tradition of rigorous evaluation using standard document corpora, predefined topics and relevance judgements can be used as a test bed for informetric analyses. We show a preliminary experiment on how both domains can be connected using the iSearch test collection, a standard information retrieval test collection derived from the open access arXiv.org preprint server. In this paper the aim is to draw a conclusion about the appropriateness of iSearch as a test bed for the evaluation of a retrieval or recommendation system that applies informetric methods to improve retrieval results for the user. Based on an interview study with physicists, bibliographic coupling and author-co-citation analysis, important authors for ten different research questions are identified. The results show that the analysed corpus includes these authors and their corresponding documents. This study is a first step towards a combination of retrieval evaluations and the evaluation of informetric analyses methods.

**Conference Topic**
Old and New Data Sources for Scientometric Studies: Coverage, Accuracy and Reliability (Topic 2), Research Fronts and Emerging Issues (Topic 4) and Open Access and Scientometrics (Topic 10).

**Introduction**
Informetric analyses are generally used to gain insights into a scientific domain and to better understand scholarly activities. A common approach is the use of statistical modelling or visualization techniques to get a more profound overview of a scientific domain or a specific topic. The process of science modelling tries to describe and formalize these approaches. Examples for methods that are used in the science modelling community are co-occurrence or co-authorship analyses or bibliographic coupling (see Scharnhorst, Börner & Besselaar, 2012). While in most cases these models are used to make scientific rankings or to draw so-called science maps, some approaches try to combine science modelling and information retrieval (IR) research (Mutschke et al., 2011).
Authors like Ingwersen (2012) propose a more application-driven view on informetrics. The main idea is that a more profound insight into a science system can be exploited to support the search process in a scholarly information system. Entities that are usually observed are authors, topics or publication organs like journals, publishers etc. The more we know about the different entities and their connection to each other in the scientific publication system, the more we can use this information to enrich the retrieval process. A classic example of this is the Bradfordizing method proposed by White (1981), where typical power-law distributions in bibliographic data sets are used to offer a different ranking mechanism. Up to then Bradford's Law was only used to detect core journals in a scientific field but this qualitative information was not used in actual retrieval systems. It was clearly shown that highly co-occurring attributes have a strong selectivity and can be applied as a ranking weight which can lead to different view on the document space (Schaer, 2011).

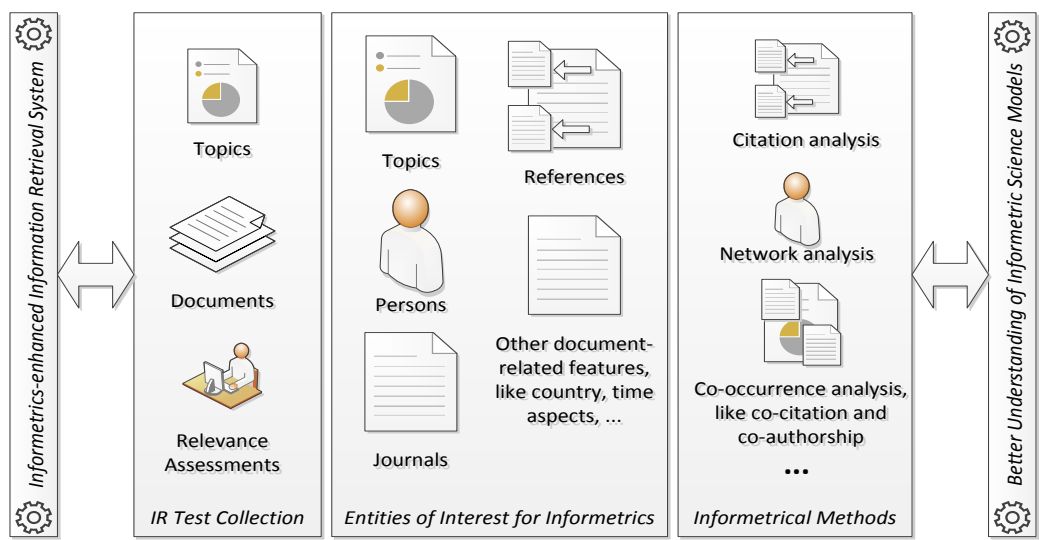

**Figure 1. Mutual benefits of IR test collections and informetric analysis methods.**

While in the IR community a decades-long evaluation tradition exists (with evaluation campaigns like TREC[1] or CLEF[2]), informetrics lacks this kind of tradition. To overcome this gap Mutschke et al. (2011) proposed the use of standard IR evaluation methods as a test-bed and "litmus test" for science models. The interconnection between IR test collections, IR evaluation methodologies and informetrics is shown in figure 1. The overall idea is an informetrics-enhanced IR system that incorporates all the previous elements to complement existing approaches through a deeper understanding of informetric science models.

The following paper describes the outcomes of a preliminary experiment on analysing a standard IR evaluation collection. We used the iSearch[3] test collection as well as data derived from an earlier experiment and an interview series with physicists. The information from the experiment is used to cross-check whether (a) the specific topics the physicists were interested in and (b) the important authors identified during the interviews are included in the iSearch corpus. If the results are positive, this standard IR corpus might serve as a basis for further retrieval and science model testing.

In the next section we will describe both data sets we want to map: the iSearch test collection and the topics and important authors extracted from social information and informetric analyses as well as interviews. Thereafter the results of the mappings are summarised and we will discuss the outcomes of this preliminary experiment in the final section.

**Data Sets**

*The iSearch Test Collection*

The iSearch test collection (Lykke et al., 2010) consists of the three standard parts of an IR test collection: (1) a corpus of documents, (2) a set of topics, and (3) relevance assessments. The corpus consists of documents from the physics domain: 18,222 monographic library records, 160,168 scientific papers and journal articles in PDF full texts and their corresponding metadata, as well as 274,749 abstracts with their corresponding metadata. Additionally the data set includes more than 3.7 million extracted internal citations. The monographic records were extracted from the Danish National Library and the full text and metadata sets were crawled from the arXiv.org open-access/preprint repository. The set of 65 topics and their relevance assessments (~200 per topic) were extracted from 23 lecturers, PhD and MSc students from three physics departments. Up to now this test collection was mainly used in the domain of contextual and task-based IR research because of the rich and realistic search tasks that allow an in-depth analysis of user intentions and expectations in the retrieval task.

---

[1] http://trec.nist.gov
[2] http://www.clef-initiative.eu
[3] http://itlab.dbit.dk/~isearch/?q=node/1

The deposition of article preprints at arXiv.org is usually best practice in most physics working groups, and so we see potential in the document corpus itself because of the size and coverage. This corpus is a rich document set from a scientific domain (in contrast to other data sets without the scientific and discipline relatedness, like the typical TREC data sets) and includes everything needed to carry out an IR evaluation (in contrast to other scientific literature corpora like INSPEC, Scopus or the Web of Science). On the other hand an author who deposits his paper is only supposed to provide a minimal set of (unstructured) metadata. In fact there are very few instructions and rules on how to enter the metadata. This results in a large but very heterogeneous document set.

*Topics and Important Authors in the Physics Domain*

To find out which resources are relevant for a user and which are not, user feedback and relevance judgements are needed. iSearch only includes such judgements for documents, but since we want to focus on important authors we need additional information. In our approach we obtained this user information from semi-structured interviews. These evaluations were part of a project to recommend collaboration partners to ten participating scientists (Heck, Peters & Stock, 2011). The aim was to recommend to a researcher authors who have similar research interests and thus could be potential collaborators. Therefore the interviewees should state whether the recommended authors are important for their current research.

The author names the physicists should evaluate were extracted using social information data. In the Web of Science[4] author names were gained using bibliographic coupling of authors, i.e. authors who have many references in common with the target researchers (the physicists) are supposed to be similar. Thus their written papers might be relevant and they might be potential collaboration partners. In Scopus[5] those authors who were co-cited many times with our target researchers were extracted (White & Griffith, 1981). In CiteULike[6] those authors were supposed to be similar whose articles have tags (assigned by the service's users) in common with the target researchers' articles, or whose articles were bookmarked by users, who also bookmarked the target scientists' papers (collaborative filtering, see e.g. Marinho et al., 2011). These author names were rated by the target scientists on a scale from 1 (not relevant for current research) to 10 (highly relevant for current research). Furthermore each physicist described his research interests with specific terms during the interview.

**Data Mapping in the iSearch Corpus**

*Design of the Experiment*

To prove the assumptions formulated in the introduction we first have to test whether the iSearch corpus is an appropriate tool to do such experiments. One criterion is that the set includes articles written by the important authors identified in the interviews. If the physicist searches for literature in his research domain, he would expect to find articles that are very relevant for his research topic. Thus the articles should be written by those authors the physicist has claimed important for his research. For the analysis of the iSearch corpus we use those authors the target physicists claimed important for their current research. We call them important authors. Important authors are those authors who in the evaluation process were rated with 5 or higher and who were explicitly named by the physicists. If the iSearch corpus includes articles from these authors – derived from methods like author-co-citation and bibliographic coupling – a retrieval system including information from science models could be evaluated on the basis of this corpus.

---

[4] http:/www.webofknowledge.com
[5] http:/www.scopus.com
[6] http://www.citeulike.org/

Table 1. Descriptions of research interests and research topics of the 10 physicists

| Topic ID | Description of research interest |
|---|---|
| sci001 | Modelling blood flow processes relating to viscosity and the formation of diseases. Analysing properties of polymers and microswimmers for medical obligations. |
| sci002 | Biomolecular multiscale simulations concerning Alzheimer disease. Analysing protein aggregation and protein-protein interaction like amyloid ß-peptide. |
| sci003 | Multiscale protein modelling and computational simulation. Analysing the properties and dynamics of fluids and polymers. |
| sci004 | Interested in polymer catalysis and neutron scattering. |
| sci005 | Analysing polymer-membrane interactions and the diffusion of red blood cells. |
| sci006 | Spintronics in carbon nanostructures, carbon nanotubes and the raman spectroscopy. |
| sci007 | Interested in photoelectron spectroscopy, (ferro) magnetic and electronic properties. |
| sci008 | Simulation of crumpled elastic sheets and its mechanical deformation. Buckling of capsid proteins. |
| sci009 | X-ray and neutron scattering in high-correlated electron systems and the building of instruments. |
| sci010 | Analysing dynamics of glass-forming liquids. Interested in inelastic neutron scattering, dielectric spectroscopy and rheology. Doing simulations of polymers and other amorphous material. |

In the interview each physicist described his research interests and research focuses with appropriate terms (see table 1). Our assumption is: If the physicist searches for literature he would probably use those terms as search terms he used to describe his research interest and research focus with. Thus the terms derived from the descriptions of the physicists' research focus (further described as topics sci001 – sci010) are used as search terms in the iSearch corpus. As some single terms would describe a research focus in a very common way, these terms are only used in combination; therefore the actual query composition was done manually based on the outcomes of the interviews to best reflect the physicists' interests. We indexed all available metadata (~453,000) and the full text data sets (~160,000) in the Solr[7] search engine and applied a standard Porter stemmer and an English stop-word list. The search is done using Solr's standard retrieval method, which is based on an extended Boolean model that allows the extraction of co-occurring entities (facets). We extracted all author names and the number of their articles from the retrieved documents. We only focused on authors and will leave the analysis of journals and references for future work.

*Results*

The physicists named 18 to 55 authors who are relevant for their current research (column 2 in table 2). We searched for those authors in the iSearch corpus. To secure correctness of the important authors and obviate author ambiguity, the author names were verified manually on the basis of co-authorship, article title and journal title. In our results we analysed two aspects, namely the important authors and their articles. We used three different data sets:
1. The whole iSearch corpus.
2. One subset per physicist, which was retrieved using the physicist's topic-describing terms by searching in title, abstract and full text (see previous section).
3. The top 50 documents of set number 2 ranked by Solr's TF*IDF implementation.

The left part of table 2 shows the coverage of the authors in the iSearch corpus, meaning at least one document of an important author can be found in the corpus. 199 of 287 unique important authors (IA) are in the corpus; on average they have nearly 19 articles in iSearch. For each topic at least 57% of IA are in the iSearch corpus. Sci006, with 29 named IA, even has 100% coverage. When using the terms describing the physicists' research interests as query terms (the subsets), nearly 70% of the previously named IA were included. But under the top 50 ranked articles there are on average only 4.8 IA. Of course the coverage depends on the time the corpus was created. Some interviewed physicists are rather novice researchers, i.e. they also named novice physicists, who are not in iSearch, as important for their research.

---
[7] http://lucene.apache.org/solr

**Table 2. Overview on the coverage of important authors (IA) and documents of important authors (IAD) within three different document pools: (1) the whole iSearch corpus, (2) a topical subset and (3) the top 50 TF*IDF ranked documents.**

| Topic | Important authors | | | | Documents of important authors | | | |
|---|---|---|---|---|---|---|---|---|
| | IA named by Researchers | IA in iSearch | IA in topic subset | IA in top 50 | Total docs in topic subset | IAD in iSearch | IAD in topic subset | IAD in top 50 |
| sci001 | 35 | 20 | 7 | 3 | 3152 | 291 | 24 | 3 |
| sci002 | 27 | 17 | 7 | 0 | 1700 | 147 | 11 | 0 |
| sci003 | 20 | 12 | 12 | 3 | 94205 | 142 | 123 | 2 |
| sci004 | 24 | 17 | 16 | 5 | 16042 | 214 | 134 | 4 |
| sci005 | 45 | 28 | 24 | 7 | 25169 | 299 | 185 | 12 |
| sci006 | 29 | 29 | 28 | 13 | 61132 | 928 | 426 | 10 |
| sci007 | 18 | 16 | 15 | 0 | 80846 | 283 | 207 | 0 |
| sci008 | 55 | 34 | 22 | 10 | 39570 | 590 | 116 | 11 |
| sci009 | 21 | 20 | 18 | 2 | 34814 | 723 | 274 | 1 |
| sci010 | 21 | 14 | 14 | 3 | 57368 | 223 | 147 | 5 |
| avg. | 29.5 | 20.7 | 16.3 | 4.8 | 41399.8 | 384 | 164.7 | 4.8 |

In the right part of table 2 we report on the coverage of the documents authored by the important authors (IAD) within the three described document pools. Note that the numbers of IAD in iSearch are approximated as it cannot be proved that every single document is really written by the correct important author. That means author ambiguity can be eliminated in the top 50 documents and for the most part in the subsets, but not in the corpus. Column 6 in table 2 (total docs in topic subset) shows the number of documents that are found with the topics showed in table 1. In these subsets the number of articles written by IA (column IAD in topic subset) ranges from just 11 documents for topic sci002 up to 426 for topic sci006 (avg. of 164.7). Within the top 50 documents on average only 4.8 IAD were included. For two topics (sci002 and sci007) no single IA or IAD were included in the top 50 documents.

Concerning both IA and IAD the coverage under the top 50 articles is weaker than in the total iSearch corpus and in the subsets. For example: Sci007 named 18 important authors. 16 of them are in the iSearch corpus. But a search with the query terms derived from the descriptions of the researcher's interests ranks no articles of IA under the top 50. Nevertheless in sci007 207 articles (IAD in topic subset) of 15 important authors (IA in topic subset) are found with the query terms determined by the physicist's research interest descriptions. Moreover, no correlation could be detected between the size of the topic subsets and the number of IA and IAD found within these subsets, i.e. you cannot state that the bigger the subset is, the more IA and IAD are included. E.g. sci003 has over 90,000 documents in the subset. All IA found in the total iSearch corpus are included in this subset, and about 86% of the IAD. But sci004, with only 16,042 documents, has similar results. Here all but one IA are included in the subset as well as 62% of the IAD. Sci008, which has more than twice as many documents as sci004, covers only about 65% of IA and about 20% of IAD. To summarize, it can be said that the coverage of IA and IAD is quite high in the iSearch corpus (nearly 70% of

IA could be found) but in the subsets and especially in the top 50 ranked documents the coverage is quite low in most cases.

**Discussion and Future Work**

We presented the outcomes of a preliminary experiment of mapping 10 specific scientific research interests onto the iSearch corpus. The statements of the physicists about authors being relevant for their current research are used as qualitative criterion to draw conclusions about the appropriateness of iSearch for evaluating informetric analyses.

Concerning the quite high coverage of important authors (nearly 70%), we assume that the iSearch corpus is appropriate for an evaluation of a retrieval approach that uses informetric methods to improve the retrieval process. In the future project we would like to use the references of the iSearch corpus and build a retrieval system that also applies bibliographic coupling and co-citation analyses to improve the results for the user. For the evaluation we would use not only the external feedback by the physicists, but also the relevance feedback of the iSearch corpus. However the physicists' feedback are beneficial because they include concrete relevance ratings of important authors, which can be used to make further statements about the retrieval results. The relevance feedback in the iSearch corpus allow statements about the relevance of articles, but not about concrete authors. Both articles and authors might be important for a user who searches for relevant research literature in a retrieval system.

Concerning the coverage of important authors in the top 50 subset, we suppose that good retrieval results should rank the articles of the important authors at very high positions. The TD*IDF ranking alone doesn't seem to be powerful enough. It is assumed that methods like co-citation analysis and bibliographic coupling will improve both document and also author retrieval. It should be tested at which stages and in which processes of a retrieval system these approaches could be applied. One idea is to re-rank the documents, which were retrieved by e.g. co-word analysis. Depending on the users' need, informetric methods may also be applied before co-word approaches and ranking. The analysis of important journals is another method to gain more relevant articles.

**References**

Heck, T., Peters, I. & Stock, W.G.(2011). Testing collaborative filtering against co-citation analysis and bibliographic coupling for academic author recommendation. *Proceedings of the 3rd ACM RecSys'11 Workshop on Recommender Systems and the Social Web* (pp. 16–23). Chicago: ACM.

Ingwersen, P. (2012). Citations and references as keys to relevance ranking in interactive IR. In *Proceedings of the 4th Information Interaction in Context Symposium* (p. 1). New York: ACM.

Lykke, M., Larsen, B., Lund, H. & Ingwersen, P. (2010). Developing a test collection for the evaluation of integrated search. In C. Gurrin, Y. He, G. Kazai, U. Kruschwitz, S. Little, T. Roelleke, S. Rüger & K. Rijsbergen (Eds.) *Advances in Information Retrieval* (pp. 627–630). Berlin/Heidelberg: Springer.

Marinho, L.B., Nanopoulos, A., Schmidt-Thieme, L., Jäschke, R., Hotho, A., Stumme, G. & Symeonidis, P. (2011): Social tagging recommender systems. In F. Ricci, L. Rokach, B. Shapira & P.B. Kantor (Eds.) *Recommender Systems Handbook* (pp. 615–644). Berlin: Springer.

Mutschke, P., Mayr, P., Schaer, P. & Sure, Y. (2011). Science models as value-added services for scholarly information systems. *Scientometrics*, 89, 1, 349–364.

Schaer, P. (2011): Using lotkaian informetrics for ranking in digital libraries. In C. Hoare & A. O'Riordan (Eds.) *Proceedings of the ASIS&T European Workshop 2011*. Cork: ASIS&T.

Scharnhorst, A., Börner, K. & Besselaar, P. van den (Eds) (2012). *Models of Science Dynamics Encounters Between Complexity Theory and Information Sciences*. Berlin: Springer.

White, H.D. (1981). "Bradfordizing" search output: how it would help online users. *Online Information Review,* 5, 1, 47–54.

White, H.D. & Griffith, B.C. (1981). Author cocitation: A literature measure of intellectual structure. *Journal of the American Society for Information Science*, 32, 3, 163–171.